\documentclass{article}

\usepackage{latexsym}
\usepackage{epsf}

\newcommand{\be}{\begin{equation}}
\newcommand{\ee}{\end{equation}}
\def\tr{{\rm tr}}
\newcommand{\Tr}{{\rm Tr}}
\def\hBox{\hat{\vphantom{A}\Box}}

\def\tfrac#1#2{{\textstyle {#1 \over #2}}}
\def\dfrac#1#2{{\displaystyle {#1 \over #2}}}
\def\dint{\displaystyle \int }

\def\tsum{\textstyle \sum }

\begin{document}

\begin{titlepage}
\flushright {\tt hep-th/0109130}\\[1cm]
\begin{center}
{\bf\Large Noncommutative ${\cal N}=2$ Supersymmetric Theories
             in Harmonic Superspace}\\[1cm]

{\large\bf I.L. Buchbinder\footnote{email:joseph@tspu.edu.ru}
$^\dag$ and I.B. Samsonov\footnote{email:samsonov@phys.tsu.ru}
$^\ddag$}\\[0.5cm]
\setcounter{footnote}{0}
{\it
 $^\dag$ Department of Theoretical Physics, Tomsk State Pedagogical
          University, Tomsk 634041, Russia \\
 $^\ddag$ Department of Quantum Field Theory, Tomsk State University,
          Tomsk 634050, Russia}\\[1cm]
{\bf Abstract}
\end{center}
\begin{quotation}
We discuss a formulation of harmonic superspace approach for noncommuative
${\cal N}=2$ supersymmetric field theories paying main attention on new
features arising because of noncommutativity. We begin with the known
notions of the harmonic superfield models and results obtained and consider
how these notions and results are modified in the noncommutative case.
The actions of basic ${\cal N}=2$
models, like hypermultiplet and ${\cal N}=2$ vector multiplet, are given
in terms of unconstrained off-shell superfields on noncommutative harmonic
superspace. We calculate the low-energy contributions to the effective actions
of these models. The background
field method in noncommutative harmonic superspace is developed and it is
applied to study the 1-loop effective action in general
noncommutative ${\cal N}=2$ model and ${\cal N}=4$ SYM theory.
\end{quotation}

\end{titlepage}
\section{Introduction}
Noncommutative field theory has recently attracted much attention due to
its profound links with modern string/brane theory \cite{Connes,Seib99}
(see also the review \cite{Nekr}). However, the noncommutative field
theory is very interesting itself, in particular, being non-local,
it can be causal and its structure can be studied by standard
method of quantum field theory.

Noncommutative field theories possess specific properties of UV- and
IR-divergencies which distinguish them from ordinary theories very
essentially. It was shown for noncommutative scalar field
models \cite{Min} that the UV-divergencies lead to additional
IR-singularities of effective action that was called the UV/IR-mixing. The
change in the divergencie structure can break the renormalizability of
theories that was pointed out for the models of matter and gauge fields in
\cite{Arefeva}. It is also interesting to note that the UV/IR-mixing in
$U(N)$ gauge theories is related only to additional $U(1)$ phase while
the $SU(N)$ part of effective action remains unchanged in low-energy
limit \cite{Arm}.

Construction of noncommutative supersymmetric models is formally realized
on the basis of a simple enough procedure. Noncommutativity usually
concerns only the bosonic coordinates of
superspace while the fermionic variables remain flat \cite{Ferrara}
(see an attempt to develop a superspace with noncommutativity
both in bosonic sector and in fermionic one \cite{Klemm}). Quantum properties
of ${\cal N}=1$ supersymmetric theories of matter and gauge fields are discussed
in many papers (see e.g. \cite{Terash,Zanon,Gomes,Petrov}). A study of extended supersymmetric
models deserves a special attention. For example, maximally extended
supersymmetric noncommutative ${\cal N}=4$ SYM is finite and has no UV/IR-mixing
\cite{Hash} what is convenient for investigating the low-energy behaviour
\cite{Zanon,Liu} and for constructing new UV-finite models \cite{Jack}.
Noncommutative ${\cal N}=2$ SYM has also been studied in many details
(see e.g. \cite{Armoni,Holl,Zanon}). We point out that all
actual calculations in noncommutative extended supersymmetric theories
have been done in terms of component fields or
using ${\cal N}=1$ superfield technique.

The aim of this paper is to develop a formulation of noncommutative
${\cal N}=2$ supersymmetric theories in harmonic superspace.
A harmonic
superspace approach \cite{harm} provides a unique possibility to formulate
${\cal N}=2$
supersymmetric models in terms of off-shell ${\cal N}=2$ superfields and
therefore allows to preserve manifest ${\cal N}=2$ supersymmetry what is
especially important in quantum theory. We describe a construction
of the basic noncommutative ${\cal N}=2$ models within the harmonic superspace approach and
investigate the new aspects stipulated by noncommutativity.

We begin with noncommutative generalization of hypermultiplet and
${\cal N}=2$ vector multiplet models and then demonstrate the basic examples
of one-loop calculations in these models and show how the effect of
UV/IR-mixing arises in harmonic supergraphs.
The effective action of Abelian
$q$-hypermultiplet interacting with gauge superfield is found to be free
of UV/IR-mixing and contributions of non-planar type. This allows us to
calculate the holomorphic effective action in this theory in a gauge
invariant form. We prove that it can be obtained from the holomorphic effective action
of commutative $q$-hypermultiplet just by the insertion of the
$\star$-product instead of usual multiplication. We consider also the
selfinteracting $q$-hypermultiplet and find that it is nonrenormalizable
because of wrong dimension of coupling constants as well as in commutative case.
The selfinteraction can
be induced by the quantum corrections from the four-point box diagrams as it
was established for the commutative $q$-hypermultiplet model \cite{IKZ}.

We investigate also the structure of low-energy effective action of
general noncommutative ${\cal N}=2$ model including the matter and gauge
field. At low energies the effective action is determined by the
holomorphic and non-holomorphic potentials in complete analogy with the
commutative case. To study these potentials we use the ${\cal N}=2$
background field method \cite{Buch2} adapted for noncommutative
theories. The holomorphic effective action is found for the
$q$-hypermultiplet model while the non-holomorphic potential us studied
for the noncommutative ${\cal N}=4$ SYM written in terms of ${\cal N}=2$
superfields. It is shown that the non-holomorphic effective action of
${\cal N}=4$ SYM consists of two parts related to the $U(N)/[U(1)]^N$ and
$[U(1)]^N$ sectors of $U(N)$ gauge group. In the commutative limit the
first part reproduces a standard non-holomorphic potential of commutative
${\cal N}=4$ $SU(N)$ SYM studied in \cite{Buch3,Buch4} while the second
one is responsible only for higher noncommutative corrections.
We consider the above results as a starting point
for detailed investigation of noncommutative
corrections in the ${\cal N}=2,4$ supersymmetric theories.

The paper is organized as follows. In the second section we consider a
generalization of harmonic superspace approach to the case of
noncommutative geometry. The classical actions of noncommutative
$q$-hypermultiplet and vector multiplet are given and corresponding Feynman
rules are written down here. The third section is devoted to the examples
of 1-loop quantum computations. In the fourth section we investigate the
low-energy effective action of general noncommutative ${\cal N}=2$ SYM
model and in ${\cal N}=4$ SYM on the basis of background field method. The
structures of noncommutative holomorphic potential of $q$-hypermultiplet and
non-holomorphic potential in ${\cal N}=4$ SYM are discussed here. Finally,
the section 5 is a summary.

\section{Classical actions and Feynman rules}
Standard harmonic superspace was introduced in the papers \cite{harm} to
develop the manifest ${\cal N}=2$ superfield formulation of ${\cal N}=2$
off-shell supersymmetric models. It is usually obtained from
conventional ${\cal N}=2$ superspace with the coordinates
$
z^M=\{x^\mu,\theta_{\alpha i},\bar\theta^i_{\dot\alpha}\}
$
by adding the harmonic variables $u^\pm_i$ which parameterize
the coset $SU(2)/U(1)$. As a result, the ${\cal N}=2$ harmonic superspace is
represented by the following coordinates
\be
z^M_A=\{x^\mu_A,\theta^+_\alpha,\bar\theta^+_{\dot\alpha},
\theta^-_\alpha,\bar\theta^-_{\dot\alpha},u^\pm_i \},
\label{e1}
\ee
where
$$
x^\mu_A=x^\mu-2i\theta^{(i}\sigma^\mu\bar\theta^{j)}u^+_iu^-_j,\ \
\theta^\pm_\alpha=\theta^i_\alpha u^\pm_i,\ \
\bar\theta^\pm_{\dot\alpha}=\bar\theta^i_{\dot\alpha}u^\pm_i.
$$
The main conceptual advantage of harmonic superspace is that the ${\cal N}=2$
vector multiplet and hypermultiplet can be described by unconstrained
superfields over the so called analytic subspace parameterized by the
coordinates
\be
\zeta_A=\{x^\mu_A,\theta^+_\alpha,\bar\theta^+_{\dot\alpha},u^\pm_i \},
\label{e2}
\ee
which transform through each other under ${\cal N}=2$
supersymmetric transformations.

In order to introduce the noncommutative geometry in ${\cal N}=2$ harmonic
superspace we perform the Moyal-Weyl deformations of bosonic
coordinates
\be
[x_\mu,x_\nu]=i\theta_{\mu\nu},
\label{e3}
\ee
where $\theta_{\mu\nu}$ is a constant antisymmetric tensor of dimension
$-2$
\footnote{We use the same symbol for the notation of fermionic variables
and noncommutative parameter. This can not lead to misunderstanding
since they appear in a completely different way and have nothing common.
}. We consider only the case when fermionic variables remain flat. A
general case of deformations of superspace is studied in ref.
\cite{Ferrara}. A natural multiplication of superfields on
noncommutative plain is given by the $\star$-product
\be
(\phi\star\psi)(x)=\phi(x)e^{\frac i2
\theta^{\mu\nu}\overleftarrow{\partial_\mu}\overrightarrow{\partial_\nu}}
\psi(x),
\label{e4}
\ee
which is associative but noncommutative.

As a rule, noncommutative field theories can be obtained from conventional
models of quantum field theory just by replacement of usual multiplication
by $\star$-product. In this section we construct the classical actions of
noncommutative hypermultiplet and ${\cal N}=2$ vector multiplet in harmonic
superspace. Some aspects of noncommutative ${\cal N}=2$ SYM model in harmonic
superspace such as solitonic solutions were considered in ref.
\cite{Belh}. In this paper we study only quantum features of these
theories. We use the notations for harmonic superspace objects accepted
in refs. \cite{harm} and \cite{Buch1,Buch2,Buch3,Buch4}.

\subsection{Selfinteracting $q$-hypermultiplet}
The classical action of selfinteracting $q$-hypermultiplet can be obtained
from the action of usual $q$-hy\-per\-mul\-ti\-plet \cite{harm} by insertion of
$\star$-product instead of usual multiplication
\be
\begin{array}{lll}
S&=&S_0+S_{int},\\
 S_0&=&\dint d\zeta^{(-4)}\breve q^+(\zeta)D^{++}q^+(\zeta),\\
S_{int}&=&\dint d\zeta^{(-4)}(\alpha\breve q^+\star\breve q^+\star q^+\star
q^++\beta \breve q^+\star q^+\star\breve q^+\star q^+).
\end{array}
\label{e5}
\ee
Note that there are two coupling constants $\alpha,\beta$ due to two types
of ordering of superfields. In commutative limit $\theta\rightarrow0$ the
two terms in the interaction (\ref{e5}) reduce to a single standard one
$\lambda(q^+)^2(\breve q^+)^2$.

Free classical action $S_0$ is not modified by the introduction of
noncommutativity, therefore, the two-point Green function has standard
form \cite{harm}
\be
G_0^{(1,1)}(1|2)=-\frac1{\Box_1}\frac{(D^+_1)^4(D^+_2)^4\delta^{12}
(z_1-z_2)}{(u^+_1u^+_2)^3}.
\label{e6}
\ee
The corresponding propagator (Fourier transformation of $G_0(1|2)$) is
given by
\be
\begin{array}{rl}
\epsfxsize=1.5cm
\epsfbox{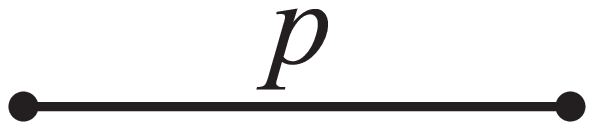}&\!\!\!\!=G_0(p)=\dfrac1{p^2}\dfrac{(D^+_1)^4(D^+_2)^4\delta^8
(\theta_1-\theta_2)}{(u^+_1u^+_2)^3}.
\end{array}
\label{e8}
\ee
The vertex can be found from the interaction $S_{int}$ as
\be
\epsfxsize=1.5cm
\epsfbox{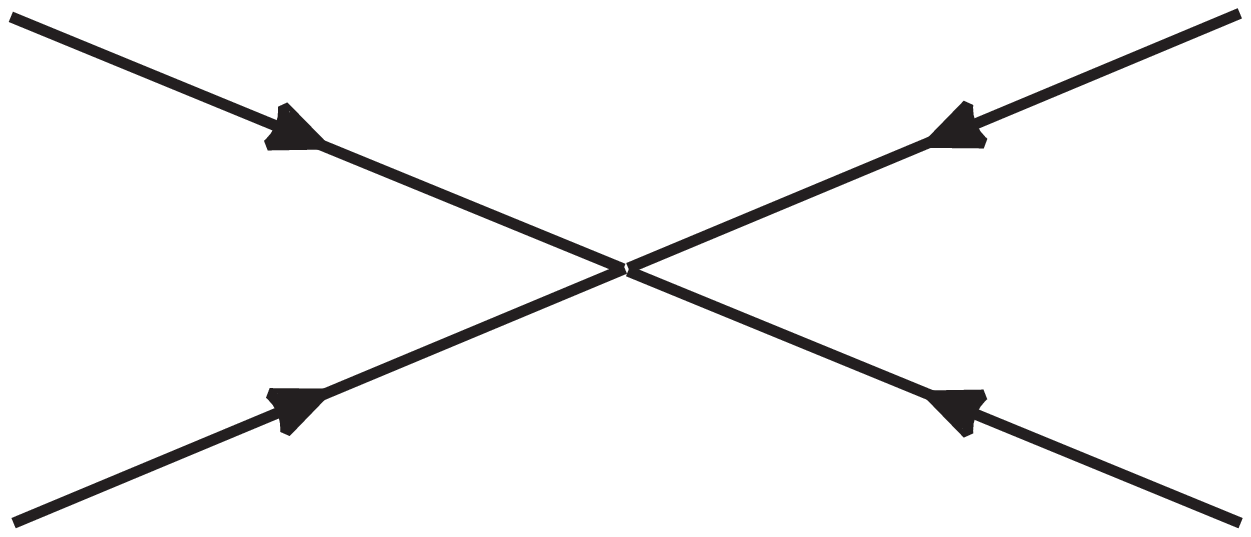}
\begin{array}{l}
=(2\pi)^4\delta^4(p_1+p_2+p_3+p_4)[\alpha \cos\frac{p_1\theta p_2}2
\cos\frac{p_3\theta p_4}2+\beta\cos(\frac{p_1\theta p_3}2+
\frac{p_2\theta p_4}2 ){}].\\
\vspace{2mm}\phantom{a}
\end{array}
\label{e9}
\ee

\vspace{-5mm}\noindent
Note that the noncommutative factor in (\ref{e9}) has the same form as in
noncommutative $\phi^4$ model studied in ref. \cite{Arefeva}.

\subsection{Noncommutative ${\cal N}=2$ SYM action}
The model of noncommutative ${\cal N}=2$ vector multiplet was considered in
refs. \cite{Zanon,Holl} using the ${\cal N}=1$ superfield approach. In terms of
${\cal N}=2$ superfields it is written as
\be
S_{SYM}=\frac1{g^2}\tr\int d^4xd^4\theta W\star W,
\label{e9.2}
\ee
where $W$ is noncommutative strength superfield defined by the equation
\be
 W=\frac i4\{\bar\nabla^{+\dot\alpha},\bar\nabla^-_{\dot\alpha} \}_\star.
\label{e9.3}
\ee
The strength $W$ can be expressed through the prepotential $V^{++}$ in the
same way as in commutative theory \cite{Zupnik87}
\be
W=\frac14\bar D^+_{\dot\alpha}\bar D^{+\dot\alpha}
 \sum\limits^\infty_{n=1}i(-ig)^{n}\int du_1\ldots du_n
 \frac{V^{++}(z,u_1)\star V^{++}(z,u_2)\star\ldots\star
 V^{++}(z,u_n)}{(u^+u^+_1)(u^+_1u^+_2)\ldots(u^+_nu^+)}.
\label{e9.4}
\ee
Note that the strength $W$ is a nonliniar function of $V^{++}$ even in
Abelian case.

Using the relation (\ref{e9.4}) the classical action of noncommutative
vector multiplet (\ref{e9.2}) can be rewritten in terms of unconstrained
vector prepotential $V^{++}$ as
\be
\begin{array}{l}
S_{SYM}=S_0+S_{int},\\
S_0=\tr\dint d^{12}z du_1du_2
\frac{V^{++}(z,u_1)V^{++}(z,u_2)}{(u^+_1u^+_2)^2},\\
S_{int}=\frac1{g^2}\tr\dint d^{12}z\sum\limits_{n=3}^\infty\frac{(-ig)^n}{n}
\dint du_1\ldots du_n\frac{V^{++}(z,u_1)\star V^{++}(z,u_2)\star
\ldots\star V^{++}(z,u_n)}{
(u^+_1u^+_2)\ldots(u^+_nu^+_1)}.
\end{array}
\label{e10}
\ee
This action is gauge-invariant under noncommutative gauge transformations
with gauge parameter $\lambda$
\be
\delta V^{++}=-D^{++}\lambda-i[V^{++},\lambda]_\star,
\label{e11}
\ee
where
\be
[V^{++},\lambda]_\star=V^{++}\star\lambda-\lambda\star V^{++}.
\label{e12}
\ee
The vector superfield $V^{++}$ and the gauge parameter $\lambda$ are the
Lie-algebra-valued analytic superfields
$$
V^{++}(\zeta)=V^{++A}(\zeta)T^A,\qquad
 \lambda(\zeta)=\lambda^A(\zeta)T^A,
$$
where $T^A$ are the generators of $U(N)$ group. Note that the interaction
in noncommutative vector superfield model (\ref{e10}) is non-polynomial even
for an Abelian gauge group.

In the simplest case it is sufficient to consider only the cubic interaction
\be
S_3=\frac{ig}3\int d^{12}zdu_1du_2du_3\frac{
V^{++}(z,u_1)[V^{++}(z,u_2), V^{++}(z,u_3)]_\star}{
(u^+_1u^+_2)(u^+_2u^+_3)(u^+_3u^+_1)},
\label{e13}
\ee
that defines the three-point vertex
\be
\epsfxsize=1.5cm
\epsfbox{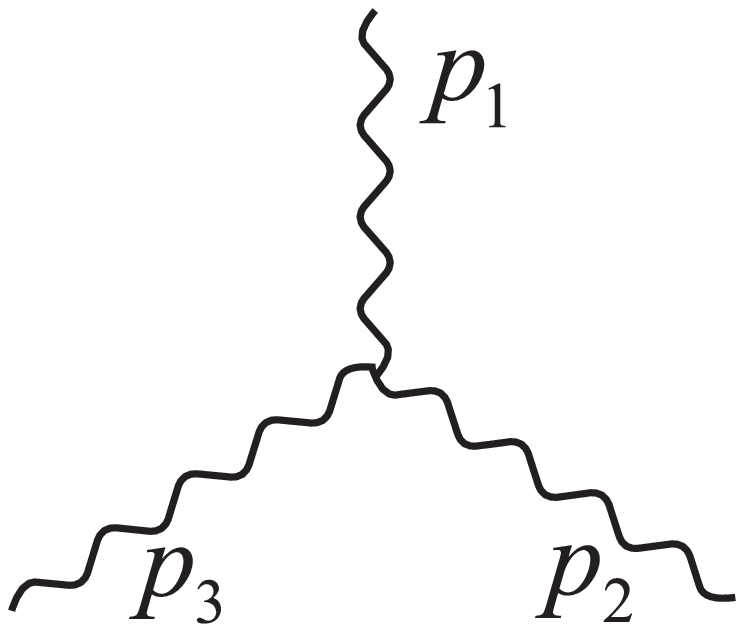}
\begin{array}l
=\frac{2g(2\pi)^4}{3(u^+_1u^+_2)(u^+_2u^+_3)(u^+_3u^+_1)}
\delta^4(p_1+p_2+p_3)\sin\frac{p_1\theta p_2}2.\\
\phantom{A}\vspace{0.9cm}\phantom{A}
\end{array}
\label{e14}
\ee

\vspace{-9mm}
The propagator of the vector superfield $V^{++}$ was found in the
works \cite{harm} (in appropriate gauge fixing) in the form
\be
\epsfxsize=1cm
\epsfbox{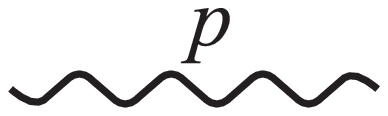}=\frac1{p^2}(D^+_1)^4\delta^8(\theta_1-\theta_2)
\delta^{(-2,2)}(u_1,u_2).
\label{e15}
\ee

In order to obtain the full action of quantum SYM theory we have to add the
ghost superfields $F$, $P$  like in \cite{harm} (with odd statistics and
$U(1)$-charge to be zero)
\be
S_{gh}=\int d\zeta^{(-4)}i\tr[F\star D^{++}(D^{++}+igV^{++})\star P].
\label{e16}
\ee
It can be shown that the ghost action (\ref{e16}) can be obtained on the
basis of standard Faddeev-Popov prescription applied to the noncommutative
vector superfield model (\ref{e10}), so the naive $\star$-product
generalization of conventional ghost action is valid here.

The corresponding Feynman rules for the ghost superfields are
\be
\begin{array}l
\epsfxsize=1.6cm
\begin{array}{rl}
\epsfbox{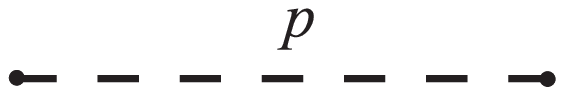}&
=-\dfrac i{p^2}(D^+_1)^4(D^+_2)^4\delta^8(\theta_1-\theta_2)
\frac{(u^-_1u^-_2)}{(u^+_1u^+_2)^3},
\end{array}
\\
\epsfxsize=1.6cm
\epsfbox{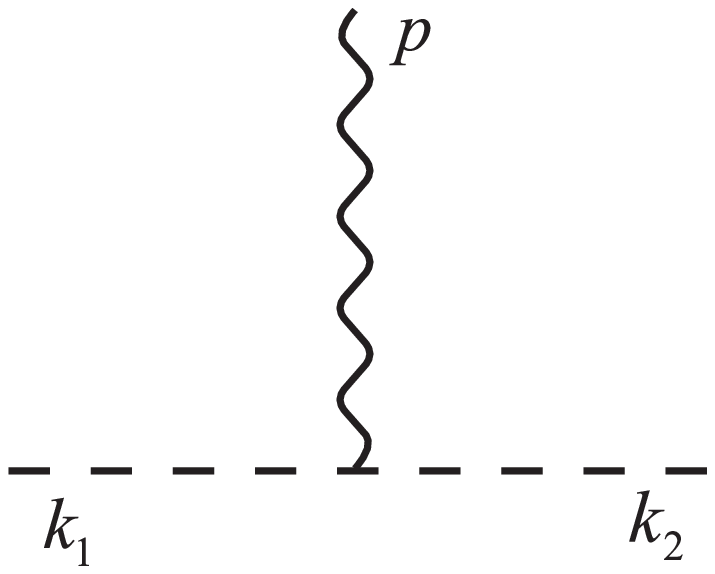}
\begin{array}l
=g(2\pi)^4D^{++}\delta^4(p+k_1+k_2)
\sin\frac{k_1\theta k_2}2.\\
\vspace{0.3cm}\phantom{A}
\end{array}
\end{array}
\label{e17}
\ee

\vspace{-0.5cm}\noindent
The sine factor in eq. (\ref{e17}) is due to odd statistics of ghosts.

\subsection{Interaction with matter}
Let us consider the massive hypermultiplet model. Mass of
$q$-hypermultiplet can be generated by central charge of ${\cal N}=2$
superalgebra \cite{Kuz97,IKZ,Buch2}. The corresponding free classical action
reads
\be
S_0=\int d\zeta^{(-4)}\breve q^+(D^{++}+iV^{++}_0)q^+,
\label{e18}
\ee
with $V_0^{++}=-\bar W_0(\theta^+)^2-W_0(\bar \theta^+)^2$,
$\bar W_0W_0=m^2$ is the mass. The propogator of this theory was found in
the refs. \cite{Zupnik86,Kuz97,IKZ} and can be represented in the form
\be
\begin{array}{rl}
G_0^{(1,1)}(1|2)=&-\dfrac1{\Box+m^2}(D^+_1)(D^+_2)^4
\left\{
\dfrac{e^{i\Omega_0(1)-i\Omega_0(2)}\delta^{12}(z_1-
z_2)}{(u^+_1u^+_2)^3}
\right\},
\end{array}
\label{e19}
\ee
or in momentum space
\be
\begin{array}{rl}
\epsfxsize=1.5cm
\epsfbox{g.eps}=&G_0^{(1,1)}(p)=\dfrac1{p^2-m^2}(D^+_1)^4(D^+_2)^4
\left\{
\dfrac{e^{i\Omega_0(1)-i\Omega_0(2)}
\delta^8(\theta_1-\theta_2)}{(u^+_1u^+_2)^3}\right\},
\end{array}
\label{e20}
\ee
where $\Omega_0=-\bar W_0\theta^+\theta^--W_0\bar\theta^+\bar\theta^-$ is
a bridge superfield in the case under consideration \cite{IKZ}.

Interaction of gauge superfields with matter can be introduced by a
minimal way. We consider two types of representations of gauge group,
fundamental and adjoint:
\be
S_{f,int}=\dint d\zeta^{(-4)}\breve Q^+\star V^{++}\star Q^+,
\label{e21}
\ee\be
S_{ad,int}=i\tr\dint d\zeta^{(-4)}\breve Q^+\star[V^{++},Q^+]_\star,
\label{e22}
\ee
where $Q^+=q^{+A}T^A$, $\breve Q^+=\breve q^{+A}T^A$, $V^{++}=V^{++A}T^A$,
$T^A$ are the generators of $U(N)$ gauge group. The commutator in
the action (\ref{e22}) is written as
\be
\begin{array}{rl}
[V^{++},Q^+]_\star=&V^{++}\star Q^+-Q^+\star V^{++}\\
=&\frac12[V^{++A},q^{+B}]_\star\{T^A,T^B\}
 +\frac12\{V^{++A},q^{+B}\}_\star[T^A,T^B]\\
=&-\frac i2[V^{++A},q^{+B}]_\star d^{ABC}T^C
 +\frac12\{V^{++A},q^{+B}\}_\star f^{ABC}T^C,
\end{array}
\label{e23}
\ee
where we have used the relations
\be
[T^A,T^B]=f^{ABC}T^C,\qquad \{T^A,T^B\}=-id^{ABC}T^C,\qquad
\tr(T^AT^B)=-\frac12\delta^{AB},
\label{e24}
\ee
$f^{ABC}$ and $d^{ABC}$ are the structure constants of $U(N)$ group
\cite{Holl}.

The vertex is found from the expressions (\ref{e21},\ref{e22}) with the
help of relations (\ref{e23},\ref{e24}) in the form
\be
\epsfxsize=1.8cm
\epsfbox{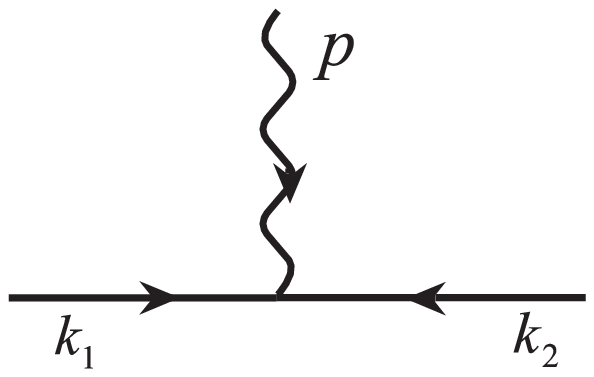}
\begin{array}{l}
=(2\pi)^4\delta^4(p+k_1+k_2)
\left\{\begin{array}l
e^{-\frac i2k_1\theta k_2},{\rm\ fundamental\ representation,}\\
F^{ABC}(k_1,k_2),{\rm\ adjoint\ representation, }
\end{array}
\right.
\\
\vspace{0.5cm}\phantom{A}
\end{array}
\label{e25}
\ee

\vspace{-0.8cm}\noindent
where $$F^{ABC}(k_1,k_2)=d^{ABC}\sin\frac12k_1\theta k_2+f^{ABC}\cos\frac12
k_1\theta k_2.$$

Now we can apply the Feynman rules obtained for noncommutative models of
hypermultiplet and vector multiplet to compute quantum corrections in
these theories.

\section{Examples of 1-loop computations}
\subsection{Noncommutative hypermultiplet four-point function}
For the first example we consider the 1-loop four-point correction to the
effective action of hypermultiplet (\ref{e5})
\be
\begin{array}{l}
{\vphantom{1}}\,\,\,\Gamma^{(1)}_4[\breve q^+,q^+]=\Gamma_s[\breve q^+,q^+]+
 \Gamma_t[\breve q^+,q^+],\\[2mm]
\epsfxsize=2.5cm
\begin{array}l
\Gamma_s[\breve q^+,q^+]=\\
\vspace{5mm}\phantom{A}
\end{array}
 \epsfbox{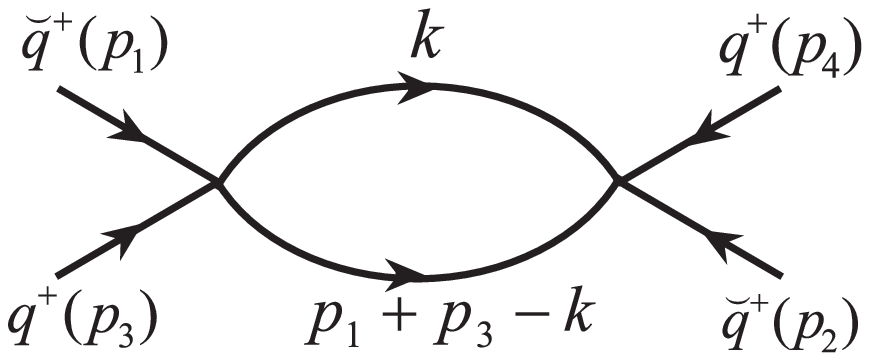},\qquad
\epsfxsize=2.5cm
\begin{array}l
\Gamma_t[\breve q^+,q^+]=\\
\vspace{5mm}\phantom{A}
\end{array}
\epsfbox{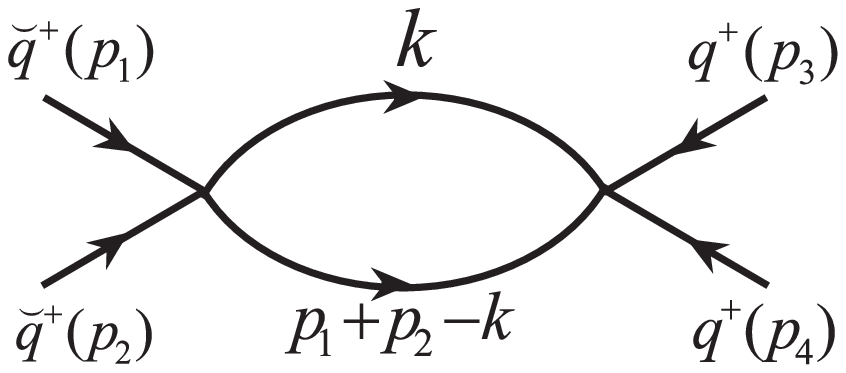}.
\end{array}
\label{e26}
\ee

\vspace{-0.6cm}\noindent
Using the Feynman rules (\ref{e8},\ref{e9}) we obtain the following
expressions for these diagrams
\be
\begin{array}{l}
\Gamma_{s,t}=\dint\tfrac{d^4p_1\ldots d^4p_4}{(2\pi)^{16}}
 d^4\theta^+_1 d^4\theta^+_2
 du_1du_2 \delta^4(p_1+p_2+p_3+p_4)\breve q^+(1)\breve
 q^+(2) q^+(3) q^+(4)\\
\qquad \times\tfrac{(D^+_1)^4(D^+_2)^4}{(u^+_1u^+_2)^3}\delta^8(\theta_1-\theta_2)
 \tfrac{(D^+_1)^4(D^+_2)^4}{(u^+_1u^+_2)^3}
\delta^8(\theta_1-\theta_2)
I_{s,t}(p_1,\ldots,p_4),
\end{array}
\label{e27}
\ee
where
\be
I_{t}(p_1,\ldots,p_4)=
2\dint d^4k\frac{F_{t}(p_1,\ldots,p_4,k)}{k^2(p_1+p_2-k)^2},\qquad
I_{s}(p_1,\ldots,p_4)=
8\dint d^4k\frac{F_{s}(p_1,\ldots,p_4,k)}{k^2(p_1+p_3-k)^2},
\label{e28}
\ee
$F_{s,t}(p_1,\ldots,p_4,k)$ are the functions the structure of which is
stipulated by noncommutativity
\be
\begin{array}{rl}
F_t(p_1,\ldots,p_4,k) =&(\alpha\cos\frac{p_1\theta p_2}2
 \cos\frac{k\theta(p_1+p_2)}2+
 \beta\cos(\frac{p_1\theta k}2+\frac{p_2\theta(p_1-k)}2))\\&
\times(\alpha\cos\frac{k\theta(p_1+p_2)}2\cos\frac{p_3\theta p_4}2+
 \beta\cos(\frac{k\theta p_3}2-\frac{(p_3+k)\theta p_4}2)),\\
F_s(p_1,\ldots,p_4,k)=&(\alpha\cos\frac{k\theta p_1}2
 \cos\frac{(p_1-k)\theta p_3}2+\beta\cos(\frac{k\theta(p_1+p_3)}2+
  \frac{p_1\theta p_3}2))\\&
\times(\alpha\cos\frac{k\theta p_4}2\cos\frac{(p_4+k)\theta p_2}2+
 \beta\cos(\frac{p_2\theta p_4}2-\frac{k\theta(p_1+p_3)}2)).
\end{array}
\label{e29}
\ee
Further computations are very similar to usual ones in commutative
$q^4$-model considered in \cite{harm}. The resulting expression for the
effective action (\ref{e27}) can be rewritten as
\be
\Gamma_{s,t}=\dint\frac{d^4p_1\ldots d^4p_4}{(2\pi)^{16}}d^8\theta
du_1du_2
\dfrac{\breve q^+(1)\breve q^+(2) q^+(3)q^+(4)
}{(u^+_1u^+_2)^2}
\delta^4(p_1+p_2+p_3+p_4)I_{s,t}(p_1,\ldots,p_4).
\label{e30}
\ee
This effective action differs from corresponding commutative one by the
presence of two functions $F_{s,t}$ in the integrals (\ref{e28}). These
functions change the UV-structure of integrals $I_{s,t}$ what lead to the
appearance of UV/IR-mixing.

To study the structure of momentum integrals (\ref{e28}) let us split the
functions $F_{s,t}$ into planar ({\it pl}) and non-planar ({\it npl}) parts
\be
F_{s,t}(p_1,\ldots,p_4,k)=F_{s,t}^{pl}(p_1,\ldots,p_4)
+F_{s,t}^{npl}(p_1,\ldots,p_4,k),
\label{e31}
\ee
where we assume that $F_{s,t}^{pl}$ do not contain the terms depending
on the momentum $k$. The planar parts are calculated explicitly
\be
\begin{array}{l}
F_t^{pl}=\frac{\alpha^2}2\cos\frac{p_1\theta p_2}
 2\cos\frac{p_3\theta p_4}2,\\
F_s^{pl}=(\frac{\alpha^2}8+\frac{\beta^2}2)\cos(\frac{p_1\theta p_3}2+
 \frac{p_2\theta p_4}2)+\frac{\alpha\beta}2\cos(-\frac{p_1\theta p_3}2+
 \frac{p_2\theta p_4}2).
\end{array}
\label{e32}
\ee
Substituting the expressions (\ref{e32}) into integrals (\ref{e28})
\be
\begin{array}{l}
I^{pl}_{t}(p_1,\ldots,p_4)=2F^{pl}_{t}(p_1,\ldots,p_4)
\dint\frac{d^4k}{k^2(p_1+p_2-k)^2},\\
I^{pl}_{s}(p_1,\ldots,p_4)=
8F^{pl}_{s}(p_1,\ldots,p_4)\dint\frac{d^4k}{k^2(p_1+p_3-k)^2},
\end{array}
\label{e33}
\ee
we see that the planar contributions have the same divergencies as
corresponding diagrams of commutative $q$-hypermultiplet. The momentum
integrals in (\ref{e33}) have the IR-divergence (at low external momentum)
which can be
avoided in massive theory and UV-divergence which can not be renormalized
since the coupling constants $\alpha,\beta$
 have the dimension $-2$ as in commutative
theory \cite{harm}. So, the selfinteracting noncommutative
$q$-hypermultiplet model is nonrenormalizable as well as the
corresponding commutative theory.

Let us study now the structure of non-planar diagrams. One can show
that all non-planar terms of functions (\ref{e29}) look like
$$
\cos\frac{p_1\theta p_2}2\cos\frac{p_3\theta p_4}2\cos(k\theta(p_1+p_2))
$$
with various combinations of external momenta. They define the structure
of momentum integrals of non-planar type
\be
\begin{array}{rl}
I^{npl}_t(p_1,\ldots,p_4)&\sim
 \cos\frac{p_1\theta p_2}2\cos\frac{p_3\theta p_4}2
\dint d^4k\frac{e^{ik\theta(p_1+p_2)}}{k^2(p_1+p_2-k)^2}\\
&=\cos\frac{p_1\theta p_2}2\cos\frac{p_3\theta p_4}2
\dint_0^1d\xi
K_0(\sqrt{\xi(1-\xi)P}),
\end{array}
\label{e34}
\ee
where
$$
P=(p_1+p_2)^2\cdot(p_1+p_2)\circ(p_1+p_2),
$$
$p_1\circ p_2=p_1^\mu(\theta\theta)_{\mu\nu}p_2^\nu$, $K_0$ is the
modified Bessel function. This expression has no UV-divergence but it is
singular at low momenta $p_i\rightarrow0$ due to the asymptotics of Bessel
function
\be
K_0(x)\stackrel{x\rightarrow0}{\longrightarrow}-\ln\frac x2+{\rm finite}.
\label{e35}
\ee
Such a singularity of effective action can not be avoided by the
introduction of mass of hypermultiplet and defines the well known
UV/IR-mixing \cite{Min}. At low momenta the leading part of momentum
integral
\be
\int d^4k\frac{e^{ik\theta(p_1+p_2)}}{k^2(p_1+p_2-k)^2}\sim
\ln\frac{1}{P}
\label{e36}
\ee
is singular in commutative limit $\theta\rightarrow0$
what is responsible for the
UV/IR-mixing. Therefore the effective action of noncommutative
hypermultiplet can not be reduced to a standard one in commutative limit.

As a result, the model of noncommutative hypermultiplet is
nonrenormalizable and has the UV/IR-mixing in the sector of non-planar
diagrams.

\subsection{1-loop two-point diagram in ${\cal N}=2$ SYM}
The next example is the 1-loop two-point function in noncommutative ${\cal N}=2$
SYM. It consists of gauge and ghost loops
\be
\epsfxsize=3cm
\begin{array}r
\Gamma^2=\\
\vspace{7mm}\phantom{A}
\end{array}
\epsfbox{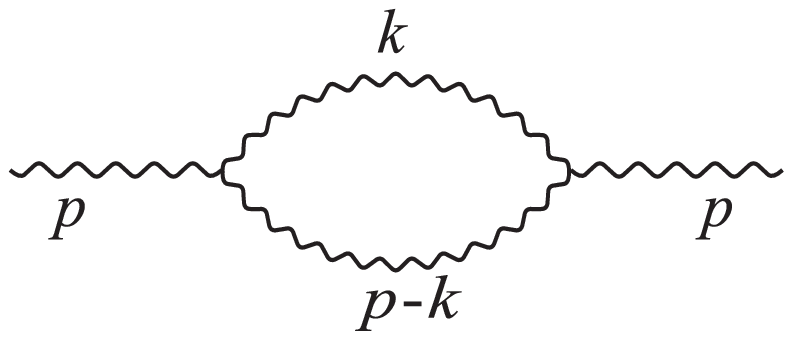}\\[-0.6cm]
\epsfxsize=3cm
\begin{array}r
\ \ \ \ +
\vspace{7mm}\phantom{A}
\end{array}
\epsfbox{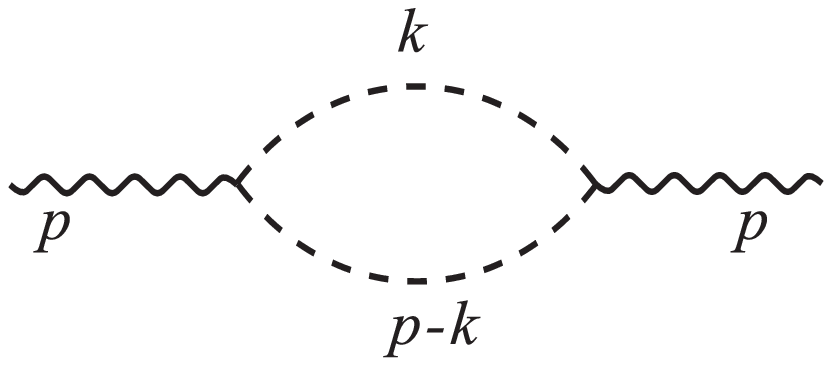}
\label{37}
\ee
which can be computed with the help of Feynman rules
(\ref{e14},\ref{e15},\ref{e17}). We skip all the calculations because they are
very similar to ones given in the previous section. The resulting
expression for the effective action is represented as a sum of planar
and nonplanar parts
$$ \Gamma_2=\Gamma_2^{pl}+\Gamma_2^{npl},
$$
\be
\Gamma_2^{pl}=\frac{-g^2}{(2\pi)^8}\int d^4pd^8\theta du_1du_2
\int\frac{d^4k}{k^2(p-k)^2}
\frac{V^{++}(p,\theta,u_1)V^{++}(-p,\theta,u_2)}{(u^+_1u^+_2)^2},
\label{e38}
\ee
\be
\begin{array}{rl}
\Gamma_2^{npl}=&\frac{g^2}{(2\pi)^8}\dint{d^4pd^8\theta du_1du_2}
\dint\frac{d^4k\cos k\theta p}{k^2(p-k)^2}
\frac{V^{++}(p,\theta,u_1)V^{++}(-p,\theta,u_2)}{(u^+_1u^+_2)^2}
\\
=&
\frac{g^2}{(2\pi)^8}\dint d^4pd^8\theta du_1du_2
\frac{V^{++}(1)V^{++}(2)}{(u^+_1u^+_2)^2}
\dint_0^1d\xi K_0(\sqrt{\xi(1-\xi)p^2\cdot p\circ p}).
\end{array}
\label{e39}
\ee
The momentum integral in the planar part of the effective action is
UV-divergent. This divergence is eliminated by standard
renormalization of coupling constant $g$. The nonplanar part
$\Gamma^{npl}$ has no UV-divergence due to the $\cos k\theta p$
factor in momentum integral which is expressed through the Bessel
function $K_0$. The Bessel function is finite at large momenta but it is
singular at small $p$ (\ref{e35}). A leading term at low energies in
$\Gamma^{npl}$ is of the form
\be
\Gamma^{npl}[V^{++}]=\frac{g^2}{2(2\pi)^8}
\int d^4pd^8\theta du_1du_2\ln\frac{1}{p^2\cdot p\circ p}
\frac{V^{++}(p,\theta,u_1)V^{++}(-p,\theta,u_2)}{(u^+_1u^+_2)^2}.
\label{e40}
\ee
This expression is singular in the limit $\theta\rightarrow0$ and it
defines the UV/IR-mixing in the ${\cal N}=2$ SYM model.

\subsection{Hypermultiplet loop}\label{hyper}
Let us consider the two-point diagram of massive q-hypermultiplet
in external Abelian vector superfield
\be
\begin{array}r
\Gamma_2[V^{++}]=\\
\vspace{0.5cm}\phantom{A}
\end{array}
\epsfxsize=2.7cm
\epsfbox{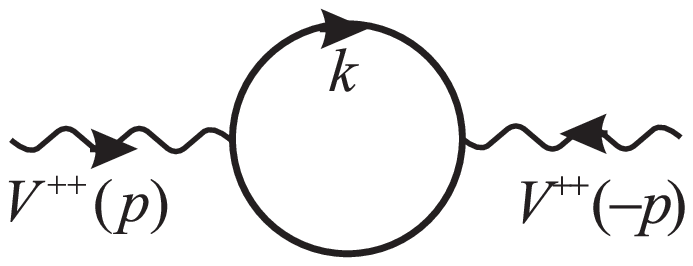},
\label{e41}
\ee

\vspace{-0.5cm}\noindent
where the Feynman rules are given by eqs. (\ref{e20},\ref{e25}).
We study both fundamental and adjoint representation of $U(1)$
gauge group.

One can show that for the fundamental representation this diagram has
no nonplanar contributions and it is equal to the corresponding diagram in
commutative theory (we omit all the calculations and give only the
final results):
\be
\Gamma_2^{fund}=\frac{2}{(2\pi)^8}\int d^4pd^8\theta du_1 du_2
\frac{V^{++}(p,\theta,u_1)V^{++}(-p,\theta,u_2)}{(u^+_1u^+_2)^2}
\int\frac{d^4k}{(k^2+m^2)((p-k)^2+m^2)}.
\label{e42}
\ee
This expression has a standard UV-divergence which has to be renormalized
as usual. There is no UV/IR-mixing since no nonplanar contributions
appear.

In the adjoint representation the two-point function (\ref{e41}) receives
both planar and nonplanar contributions:
\be
\begin{array}l
\Gamma^{ad}_2=\Gamma_{pl}+\Gamma_{npl}
\\
\Gamma_{pl}=\dint\frac{d^4p}{(2\pi)^8}d^8\theta du_1du_2
\frac{V^{++}(1)V^{++}(2)}{(u^+_1,u^+_2)^2}
\dint \frac{d^4k}{(k^2+m^2)((p-k)^2+m^2)},
\\
\Gamma_{npl}=-\dint\frac{d^4p}{(2\pi)^8}d^8\theta du_1du_2
\frac{V^{++}(1)V^{++}(2)}{(u^+_1,u^+_2)^2}
\dint_0^1 d\xi K_0(\sqrt{\xi(1-\xi)m^2p\circ p}).
\end{array}
\label{e43}
\ee
Using the asymptotics of Bessel function (\ref{e35}) one can find that
the leading low-energy contribution is of the form
\be
\Gamma^{npl}[V^{++}]=-\frac12
\int\frac{d^4p}{(2\pi)^8} d^8\theta du_1du_2\ln\frac{1}{m^2 p\circ p}
\frac{V^{++}(p,\theta,u_1)V^{++}(-p,\theta,u_2)}{(u^+_1u^+_2)^2}.
\label{e44}
\ee
The equation (\ref{e44}) is responsible for the UV/IR-mixing in this
diagram.

\subsection{Induced hypermultiplet selfinteraction}
The last example is the four-point box diagram with hypermultiplet
external lines of the type
\be
\begin{array}c
\Gamma_4=\Gamma_s+\Gamma_t,\\[1mm]
\begin{array}r
\Gamma_s=\\
\vspace{1cm}\vphantom{A}
\end{array}
\epsfxsize=2.5cm
\epsfbox{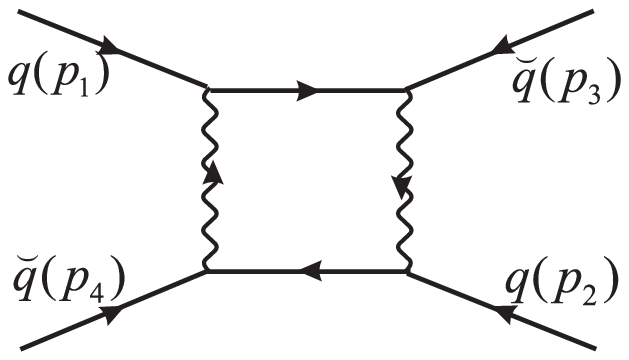},\qquad
\begin{array}r
\Gamma_t=
\vspace{1.3cm}\vphantom{A}
\end{array}
\epsfxsize=2.5cm
\epsfbox{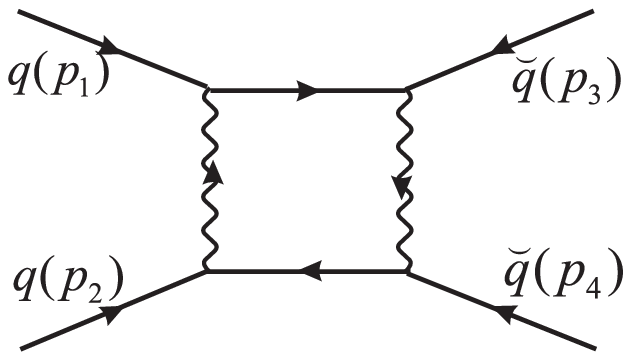}.\\[-0.2cm]
\end{array}
\label{e45}
\ee

\vspace{-0.5cm}\noindent
In the low energy approximation the loops shrink down to points and the
finite parts of these diagrams give the coupling constants $\alpha,\beta$
of selfinteracting $q$-hypermultiplet (\ref{e9}) induced only by quantum
corrections. Such an induced selfinteraction in the commutative hypermultiplet
model was considered in \cite{IKZ}.

To compute these diagrams we apply the Feynman rules obtained
(\ref{e15},\ref{e20},\ref{e25}) and, after simple transformations, we
obtain the following expressions
\be
\begin{array}l
\Gamma_{s,t}[\breve q^+,q^+]=
 -\frac{ig^4}{(2\pi)^{16}}\dint d^4p_1d^4p_2d^4p_3d^4p_4
F_{s,t}(p_1,\ldots,p_4)I_{s,t}(p_1,\ldots,p_4)\\
\qquad\times\dint d^8\theta\frac{du_1du_2}{(u^+_1u^+_2)^2}q^+(p_1)q^+(p_2)
\breve q^+(p_3)\breve q^+(p_4)
e^{i\Omega_0(2)-i\Omega_0(1)}e^{i\Omega_0(2)-i\Omega_0(1)},
\end{array}
\label{e46}
\ee
where
\be
\begin{array}l
I_s(p_1,\ldots,p_4)=\frac1{(2\pi)^4}
\dint \frac{d^4k}{k^2(k+p_1+p_3)^2}
\frac1{((k+p_1)^2+m^2)((k-p_4)^2+m^2)},
\\
I_t(p_1,\ldots,p_4)=\frac1{(2\pi)^2}
\dint \frac{d^4ke^{ik\theta(p_1+p_3)}}{
(k+p_3)^2(k-p_1)^2}
\frac1{(k^2+m^2)((k+p_3+p_4)^2+m^2)},
\end{array}
\label{e47}
\ee
\be
\begin{array}l
F_s(p_1,\ldots,p_4)=e^{-\frac i2p_1\theta p_3}e^{-\frac i2p_2\theta p_4},\\
F_t(p_1,\ldots,p_4)=e^{-\frac i2p_2\theta p_1}e^{-\frac i2p_3\theta p_4}.
\end{array}
\label{e48}
\ee
It should be noted that the momentum integrals (\ref{e47}) are UV-finite.
Therefore there is no UV/IR-mixing and these diagrams are smooth in the
limits $\theta\rightarrow0$ and $p_i\rightarrow0$. At zero external
momenta both these integrals are computed exactly:
\be
I_{s,t}=\frac1{(4\pi)^2m^2}\left(
\frac1{m^2}\ln(1+\frac{m^2}{\Lambda^2_{s,t}})-\frac1{\Lambda^2_{s,t}+m^2}
\right),
\label{e49}
\ee
where $\Lambda_{s,t}$ are the parameters of infrared cutoff
which do not coincide generally speaking. The values of
$\Lambda_{s,t}$ should be taken less then the mass of the lightest particle what
corresponds to the Wilsonian low-energy effective action.

In order to extract the relevant low-energy contribution from eq.
(\ref{e46}) it is necessary to employ the covariant derivatives algebra
\cite{harm}
\be
[{\cal D}^{++},{\cal D}^{--}]=D^0,
\label{e50}
\ee
where
\be
{\cal D}^{\pm\pm}=D^{\pm\pm}+iV_0^{\pm\pm},\qquad
 V_0^{\pm\pm}=D^{\pm\pm}\Omega_0.
\label{e51}
\ee
The relation (\ref{e50}) is used to prove the identity
\be
\begin{array}l
q^+\star q^+\star\breve q^+\star\breve q^+
 e^{i\Omega_0(2)-i\Omega_0(1)} e^{i\Omega_0(2)-i\Omega_0(1)}\\
 =
\frac12 [{\cal D}_1^{++},{\cal D}_1^{--}]
 q^+\star q^+\star\breve q^+\star\breve q^+
 e^{i\Omega_0(2)-i\Omega_0(1)} e^{i\Omega_0(2)-i\Omega_0(1)}.
\end{array}
\label{e52}
\ee
Then it can be shown that only the first term
$\sim {\cal D}^{++}{\cal D}^{--}$ in this identity gives the leading
low-energy contribution. Integrating by parts one can cancel the harmonic
distribution $\frac1{(u^+_1u^+_2)^2}$ by the use of another identity
\cite{harm}
\be
D^{++}_1\frac1{(u^+_1u^+_2)^2}=
 D^{--}_1\delta^{(2,-2)}(u_1,u_2).
\label{e53}
\ee
The harmonic $\delta$-function on the right hand side of eq. (\ref{e53})
removes one of the two remaining harmonic integrals and allows us to
rewrite eq. (\ref{e46}) to the form
\be
\Gamma_{s,t}[\breve q^+,q^+]=-ig^4 I_{s,t}\cdot S_{s,t}[\breve q^+,q^+],
\label{e54}
\ee
where
\be
\begin{array}l
S_s=\dint d^4xd^8\theta du {\cal D}^{--}{\cal D}^{--}
  q^+\star\breve q^+\star q^+\star\breve q^+,\\
S_t=\dint d^4xd^8\theta du {\cal D}^{--}{\cal D}^{--}
  q^+\star q^+\star\breve q^+\star\breve q^+.
\end{array}
\label{e55}
\ee
The integrals over full ${\cal N}=2$ superspace in eqs. (\ref{e55}) can be
transformed to ones over the analytic subspace
\be
\begin{array}l
S_s=-2m^2\dint d\zeta^{(-4)} q^+\star\breve q^+\star q^+\star\breve q^+,\\
S_t=-2m^2\dint d\zeta^{(-4)} q^+\star q^+\star\breve q^+\star\breve q^+,
\end{array}
\label{e56}
\ee
where we have used the explicit form for $v^{--}$ given by eq. (\ref{e51})
and relation $\bar WW=m^2$.
As a result, we find the low-energy part of the effective action
(\ref{e54}) in the form of four-point $q$-hypermultiplet interaction
(\ref{e5})
\be
\Gamma_4=\int d\zeta^{(-4)}(\lambda_t
  \breve q^+\star\breve q^+\star q^+\star q^++
\lambda_s\breve q^+\star q^+\star\breve q^+\star q^+),
\label{e57}
\ee
where the coupling constants $\lambda_{s,t}$ appear due to quantum
corrections (\ref{e49}) as
\be
\lambda_{s,t}=\frac{g^2}{(4\pi)^2}\left(
\frac1{m^2}\ln(1+\frac{m^2}{\Lambda_{s,t}})-\frac1{\Lambda_{s,t}^2+m^2}
\right).
\label{e58}
\ee

Note that the effective action (\ref{e57}) is smooth in the limit
$\theta\rightarrow0$ and both induced coupling constants $\lambda_{s,t}$
(\ref{e58}) reduce to a single coupling constant obtained in ref. \cite{IKZ}
for the commutative hypermultiplet model.

\section{Structure of low-energy effective action of general
noncommutative ${\cal N}=2$ theory}
The total ${\cal N}=2$ theory consists of the actions of vector superfield
(\ref{e9.2}) and hypermultiplets in fundamental and adjoint
representations (\ref{e21},\ref{e22})
\begin{eqnarray}
S&=&S_{SYM}+S^f_{HP}+S_{HP}^{ad},
\label{e59}\\
S_{SYM}&=&\frac1{g^2}\tr\dint d^4x d^4\theta W\star W,
\label{e59.1}
\\
S_{HP}^f&=&\dint d\zeta^{(-4)}[\breve q^+D^{++}q^++
 ig\breve q^+\star V^{++}\star q^+],
\label{e59.2}
\\
S_{HP}^{ad}&=&\tr\dint d\zeta^{(-4)}(\breve Q^+D^{++}Q^++
 ig\breve Q^+\star[V^{++},Q^+]_\star).
\label{e59.3}
\end{eqnarray}
The action (\ref{e59}) is gauge invariant with respect to the following
gauge transformations
\be
\begin{array}l
\delta q^+=i\lambda\star q^+,\qquad
  \delta\breve q^+=-i\breve q^+\star \lambda,\\
\delta Q^+=i[\lambda,Q^+]_\star,\qquad
  \delta\breve Q^+=-i[\breve Q^+,\lambda]_\star,\\
\delta V^{++}=-D^{++}\lambda+i[\lambda,V^{++}]_\star.
\end{array}
\label{e60}
\ee

We are interested only in the part of effective action depending on the
gauge superfields but not hypermultiplets,
therefore in the most general form it reads
\be
\begin{array}{rl}
\Gamma[W,\bar W]=&\left(
\dint d^4x d^4\theta {\cal F}(W)+c.c.\right)
+\dint d^4x d^8\theta {\cal H}(W,\bar W)\\&
+{\rm\ terms\ depending\ on\ covariant\ derivatives\ of\ }W,\bar W,
\end{array}
\label{e61}
\ee
where, as usual, the complex chiral superfield $\cal F$ is called a holomorphic
potential and the real superfield $\cal H$ is a non-holomorphic potential.
If we suppose that $\cal F$ and $\cal H$ are smooth functions of strengths
and they do not contain any derivatives of $W$, $\bar W$ (as in
commutative case) then these functions do not receive any corrections
depending on the parameter of noncommutativity $\theta$. However, we loose
the manifest gauge invariance in this case since the gauge invariance
require the $\star$-product multiplication of superfields what involves
the spatial derivatives. To provide the manifest gauge invariance of
low-energy effective action one should take into account the
noncommutative corrections to $\cal F$ and $\cal H$ (corrections depending
on $\theta$). In general, the problem of finding the noncommutative analogs
of holomorphic and non-holomorphic potentials is open so far and
the question of gauge invariance of effective action in the ${\cal N}=2,4$
noncommutative SYM is still under discussion \cite{Arm,Armoni,Holl,Zanon,Liu}. But in some
specific cases (the model of $q$-hypermultiplet in fundamental
representation) the holomorphic prepotential is written in a manifestly
supersymmetric and gauge invariant form what will be shown in this section.

The structure of functions $\cal F$ and $\cal H$ in (\ref{e61})
and all quantum contributions to them can be defined on
the basis of perturbative quantum computations. The most natural tool for
this is the background field method which we consider below.

\subsection{The background field method for ${\cal N}=2$ noncommutative SYM}
The background field method for the commutative ${\cal N}=2$ SYM theory in
harmonic superspace was developed in the paper \cite{Buch1}. In this
section we generalize these results to noncommutative SYM model
(\ref{e59}).

Let us start with the classical action of noncommutative vector multiplet
(\ref{e10})
\be
S_{SYM}[V^{++}]=\frac1{g^2}\tr \int d^{12}z\sum\limits_{n=2}^\infty\frac{(-i)^n}{n}
\int du_1\ldots du_n\frac{V^{++}(z,u_1)\star V^{++}(z,u_2)\star\ldots
 \star V^{++}(z,u_n)}{(u^+_1u^+_2)(u^+_2u^+_3)\ldots(u^+_nu^+_1)}
\label{e62}
\ee
and perform a splitting of gauge superfield into background $V^{++}$ and
quantum $v^{++}$ parts
\be
V^{++}\longrightarrow V^{++}+gv^{++}.
\label{e63}
\ee
There are two types of gauge transformations:\\
i) background
\be
\delta V^{++}=-D^{++}\lambda-i[V^{++},\lambda]_\star=
 -\nabla^{++}\star\lambda,\qquad
 \delta v^{++}=i[\lambda,v^{++}]_\star;
\label{e64}
\ee
ii) quantum transformations
\be
\delta V^{++}=0,\qquad \delta v^{++}=-\frac1g\nabla^{++}\star\lambda-
 i[v^{++},\lambda]_\star.
\label{e65}
\ee
It can be shown that upon the splitting
(\ref{e63}) the classical action (\ref{e62}) can be rewritten in the form
\be
S_{SYM}[V^{++}+gv^{++}]=S_{SYM}[V^{++}]+
 \frac1{4g}\tr\dint d\zeta^{(-4)}du\, v^{++}
  \bar D^+_{\dot\alpha}\bar D^{+\dot\alpha}\bar W_\lambda
+\Delta S_{SYM}[v^{++},V^{++}],
\label{e66}
\ee
where
\be
\begin{array}l
\Delta S_{SYM}[v^{++},V^{++}]\\\ \ \ = -\tr\dint d^{12}z\sum\limits_{n=2}^\infty
 \frac{(-ig)^{n-2}}{n}\int du_1du_2\ldots du_n
 \dfrac{v^{++}_\tau(z,u_1)\star v^{++}_\tau(z,u_2)\star\ldots
 \star v^{++}_\tau(z,u_n)}{(u^+_1u^+_2)(u^+_2u^+_3)\ldots(u^+_nu^+_1)}.
\end{array}
\label{e67}
\ee
Here we introduced the notations
\be
v^{++}_\tau=e^{-i\Omega}_\star\star v\star e^{i\Omega}_\star,\quad
 W_\lambda=e^{i\Omega}_\star\star W\star e^{-i\Omega}_\star,
\label{e68}
\ee
where $\Omega$ is a "bridge" superfield corresponding to the background
field $V^{++}$ \cite{Buch1}. The classical action (\ref{e66}) is
manifestly gauge invariant with respect to the background gauge
transformations (\ref{e64}).

To fix the gauge degrees of freedom in the effective action we follow a
Faddeev-Popov anzatz. The gauge-fixing condition can be chosen in the same
form as in commutative SYM theory
\be
D^{++}v^{++}_\tau=0.
\label{e69}
\ee
Following the same steps as in commutative case \cite{Buch1}, one
obtains that the ghost and gauge-fixing actions look like this
\be
\begin{array}l
S_{FP}=\tr\dint d\zeta^{(-4)}{\bf b}\star\nabla^{++}\star
 (\nabla^{++}\star {\bf c}+ig[v^{++},{\bf c}]_\star),\\
S_{fg}=\frac1{2\alpha}\tr\dint d^{12}z du_1du_2
 \frac{v^{++}_\tau(1)v^{++}_\tau(2)}{(u^+_1u^+_2)}-
  \frac1{4\alpha}\tr\dint d^{12}zdu v^{++}_\tau(D^{--})^2v^{++}_\tau,
\end{array}
\label{e70}
\ee
where ${\bf b},\bf c$ are the ghosts, $\alpha$ is an arbitrary parameter,
we set $\alpha=-1$. Taking all these expressions
(\ref{e66},\ref{e67},\ref{e70}) together we obtain the following
expression for the effective action
\be
e^{i\Gamma_{SYM}[V^{++}]}=e^{iS_{SYM}[V^{++}]}
 \int{\cal D}v^{++}{\cal D\bf b\cal D\bf c\cal D}\phi{\cal D}\chi^{(4)}
 {\cal D}\sigma e^{iS_q[v^{++},{\bf b},{\bf c},\phi,\chi^{(4)},\sigma]},
\label{e71}
\ee
where
\be
\begin{array}l
S_q[v^{++},{\bf b},{\bf c},\phi,\chi^{(4)},\sigma]=S_2+S_{int},\\
S_2=
 -\tfrac12\tr\dint d\zeta^{(-4)}v^{++}\star\hBox_\star\star v^{++}
+\tfrac12\tr\dint d\zeta^{(-4)}\chi^{(4)}\star\hBox_\star\star\sigma  \\
\qquad\ +\tr\dint d\zeta^{(-4)}{\bf b}\star\nabla^{++}\star\nabla^{++}
  \star{\bf c}
 +\tfrac12\tr\dint d\zeta^{(-4)}\phi\star\nabla^{++}\star\nabla^{++}
  \star\phi,\\
S_{int}=-\tr\dint d^{12}z\sum\limits^\infty_{n=3}du_1du_2\ldots du_n
 \frac{(-ig)^{n-2}}{n}\frac{v^{++}_\tau(z,u_1)\star\ldots\star
  v^{++}_\tau(z,u_n)}{(u^+_1u^+_2)\ldots(u^+_nu^+_1)}\\
\qquad\ \ \;+ig\tr\dint d\zeta^{(-4)}du{\bf b}\star\nabla^{++}\star[v^{++},
 {\bf c}]_\star.
\end{array}
\label{e72}
\ee
Here $\phi$ is a Nielsen-Kallosh ghost, $\chi^{(4)}$ and $\sigma$ are the
auxiliary superfields which are necessary for correct counting of degrees
of freedom \cite{Buch1}. The action $S_2$ is defined by the operator
\be
\begin{array}c
\hBox_\star=-\frac12(D^+)^4\nabla^{--}\star\nabla^{--}
={\cal D}^m\star{\cal D}_m+\frac12\{\bar W,W\}_\star
+\frac i2(D^{+\alpha}W)\star{\cal D}^-_\alpha\\
 +\frac i2(\bar D^+_{\dot\alpha}\bar W)\star\bar{\cal D}^{-\dot\alpha}
 -\frac i2(D^{+\alpha}D^+_\alpha W)\star{\cal D}^{--}
 +\frac i8[D^{+\alpha},{\cal D}^-_\alpha]\star W
\end{array}
\label{e73}
\ee
which is a noncommutative analog of covariant analytic d'Ala\-mber\-ti\-an
\cite{Buch1}.

The quadratic part of the action (\ref{e72}) defines the structure of
1-loop effective action as
\be
\Gamma^{(1)}[V^{++}]=\frac i2\Tr_{(2,2)}\ln\hBox_\star-
 \frac i2\Tr_{(4,0)}\ln\hBox_\star
+i\Tr_q\ln\nabla^{++}_\star-\frac i2\Tr_{ad}\ln(\nabla^{++}_\star)^2,
\label{e74}
\ee
where the formal expressions of $\Tr\ln$ in eq. (\ref{e74}) are given by
the following functional integrals
\be
\begin{array}l
({\rm Det}_{(2,2)}\hBox_\star)^{-1}=
 \dint{\cal D}v^{++}{\cal D}u^{++}\exp\{
    -i\tr\dint d\zeta^{(-4)}
   v^{++}\star\hBox_\star\star u^{++}\},\\
({\rm Det}_{(4,0)}\hBox_\star)^{-1}=
 \dint{\cal D}\rho^{(4)}{\cal D}\sigma\exp\{
 -i\tr\int d\zeta^{(-4)}
   \rho^{(4)}\star\hBox_\star\star\sigma \},\\
({\rm Det}_q \nabla^{++}_\star)^{-1}=
 \dint{\cal D}\breve q^+{\cal D}q^+\exp\{
  -i\tr\int d\zeta^{(-4)}
   \breve q^+\star\nabla^{++}\star q^+\}.
\end{array}
\label{e75}
\ee
The terms in the first line of eq. (\ref{e74}) come from the actions of
vector superfield $v^{++}$ and auxiliary superfields $\chi^{(4)},\sigma$.
The term $\Tr_{ad}\ln(\nabla^{++}_\star)^2$ corresponds to the
contributions from ghosts ${\bf b},{\bf c},\phi$. The third term in eq.
(\ref{e74}) is responsible for the contributions from matter superfields
(\ref{e59.2},\ref{e59.3}).

The formal expression (\ref{e74}) determines the low-energy structure of ${\cal N}=2$
SYM model (\ref{e59}) and provides a base for perturbative quantum
computations in manifestly supersymmetric and gauge invariant way in
noncommutative theory.

The equation (\ref{e74}) is convenient for studying the structure of
effective action of noncommutative ${\cal N}=4$ SYM. The classical action of
${\cal N}=4$ SYM model written in terms of ${\cal N}=2$ superfields in harmonic
superspace reads (see \cite{Buch3,Buch4} for commutative case)
\be
S_{{\cal N}=4}[V^{++},q^+,\breve q^+]=\frac1{g^2}\tr\int d^4xd^4\theta
W^2-
 \frac1{g^2}\tr\int d\zeta^{(-4)}q^{+i}\star\nabla^{++}\star q^+_i,
\label{e76}
\ee
where
$
q^+_i=(q^+,\breve q^+),\ q^{+i}=\varepsilon^{ij}q^+_j=
 (\breve q^+,-q^+)
$
are the hypermultiplets in adjoint representation. It can be shown that
the action (\ref{e76}) possess two more hidden supersymmetries and
therefore corresponds to the ${\cal N}=4$ SYM model.

For the ${\cal N}=4$ SYM theory (\ref{e76}) the terms in the second line of eq.
(\ref{e74}) cancel each other and the 1-loop effective action reads
\be
\Gamma^{(1)}_{{\cal N}=4}[V^{++}]=\frac i2\Tr_{(2,2)}\ln\hBox_\star-
 \frac i2\Tr_{(4,0)}\ln\hBox_\star.
\label{e77}
\ee
The perturbative expansion of $\Tr\ln$
of noncommutative d'Alambertian $\hBox_\star$ in eq.
(\ref{e77}) defines the low-energy effective action of noncommutative
${\cal N}=4$ SYM which is given by non-holomorphic potential (\ref{e61}). The
structure of non-holomorphic potential in this theory will be discussed
further.

\subsection{Holomorphic potential of $q$-hypermultiplet}
Let us consider the model of massive single $q$-hy\-per\-mul\-ti\-plet in
fundamental representation of $U(1)$ gauge group with the classical action
\be
S=\int d\zeta^{(-4)}\breve q^+\star(D^{++}+iV^{++}_0+iV^{++})\star q^+
=\int d\zeta^{(-4)}\breve q^+\star\nabla^{++}\star q^+,
\label{e78}
\ee
where $V^{++}_0=-\bar W_0(\theta^+)^2-W_0(\bar\theta^+)^2$,
$W_0\bar W_0=m^2$. The propagator and the vertex of the model are given by
eqs. (\ref{e20},\ref{e25}).

The 1-loop effective action of the model can be formally written as
\be
\begin{array}c
\Gamma^{(1)}[V^{++}]=i\tr\ln(\nabla^{++}\star)=
 \sum\limits_{n=1}^\infty\Gamma_n,\\
\Gamma_n=i\frac{(-i)^n}{n}\dint d\zeta_1^{(-4)}\ldots d\zeta_n^{(-4)}
G_0(\zeta_1,\zeta_2)\star V^{++}(\zeta_2)\ldots
G_0(\zeta_n,\zeta_1)\star V^{++}(\zeta_1).
\end{array}
\label{e79}
\ee
Two-point function was calculated in the section \ref{hyper}, where it was
found that this model has no UV/IR-mixing. Now we compute the
contributions to the holomorphic effective action from arbitrary $n$-point
function with noncommutative corrections. In momentum space the $n$-point
function $\Gamma_n$ reads
\be
\begin{array}l
\Gamma_n=\frac{i(-i)^n}{n(2\pi)^{4n}}\int d^4p_1
 d^4\theta^+_1du_1
\ldots d^4p_nd^4\theta^+_ndu_n
\int d^4k\delta^4(\sum p_i)e^{-\frac i2k\theta(\sum p_i)}\\
\quad\times G_0(k+p_2+p_3+\ldots+p_n)G_0(k+p_3+\ldots+p_n)\ldots G_0(k+p_n)
 G_0(k)\\
\quad\times e^{-\frac i2\sum^n_{i<j}
 p_i\theta p_j}V^{++}(1)\ldots V^{++}(n).
\end{array}
\label{e80}
\ee
The first exponent in eq. (\ref{e80}) reduces to unity due to the relation
$$
e^{-\frac i2k\theta(\sum p_i)}\delta^4(\sum p_i)=\delta^4(\sum p_i).
$$
This equation guarantees the absence of nonplanar diagrams and
UV/IR-mixing. Therefore, the limit $\theta\rightarrow0$ is smooth and in
low-energy approximation (when we neglect all derivatives of $V^{++}$)
the factor $e^{-\frac i2\sum_{i<j}^np_i\theta p_j}$ can be dropped. As a
result, we obtain usual expression for the holomorphic effective action of
commutative $q$-hypermultiplet, calculated in the ref. \cite{Buch2}
\be
\Gamma[V^{++}]=-\frac1{64\pi^2}\int d^4xd^4\theta W^2\ln\frac{W^2}{
\Lambda^2}+c.c.
\label{e81}
\ee
The absence of $\star$-product in eq. (\ref{e81}) is evident since we
consider an approximation when $\star$-product reduces to ordinary
multiplication. A non-trivial result here is the absence of non-planar diagrams
and UV/IR-mixing.

Now consider the next order in the approximation when all derivatives
related to noncommutativity are kept. In this case we can not drop the
factor $e^{-\frac i2\sum_{i<j}^np_i\theta p_j}$ in eq. (\ref{e80}) but
use it to restore the $\star$-product of superfields due to the
identity
\be
\begin{array}l
\dint d^4p_1\ldots d^4p_n
e^{-\frac i2\sum_{i<j}^np_i\theta p_j}
 V_1(p_1)\ldots V^{++}_n(p_n) \delta^4(\sum p_i)\\=
\dint d^4xV_1^{++}(x)\star V_2^{++}(x)\star\ldots\star V^{++}_n(x).
\end{array}
\label{rel}
\ee
It should be noted that we have to express the resulting effective action
in terms of strength superfield (\ref{e9.4}) which is a
nonlinear function of $V^{++}$ even in the Abelian case. To
simplify this problem we choose the external gauge superfield in
the form
\be
\tilde W=-\frac14\int du\bar D^-_{\dot\alpha}
 \bar D^{-\dot\alpha}\tilde V^{++}(z,u),
\label{e82}
\ee
what corresponds to the first term in series (\ref{e9.4}).
With such a special background $\tilde W$ one obtains the standard expression
for holomorphic effective action \cite{Buch2}
written in terms of strength superfields $\tilde W$ where conventional
multiplication should be replaced by $\star$-product due to the relation
(\ref{rel}).
To return to arbitrary background it is necessary to restore the full
strength $W$ from $\tilde W$ with the help of gauge transformations
$$
W=e^{i\lambda}_\star\star\tilde W\star e^{-i\lambda}_\star
$$
with a special $\lambda=\lambda(W)$. Therefore, for arbitrary $W$
the holomorphic effective action of noncommutative
$q$-hypermultiplet reads
\be
\Gamma[V^{++}]=-\frac1{64\pi^2}\int d^4xd^4\theta
W\star W\star\ln_\star\frac{W}{\Lambda}+c.c.
\label{e84}
\ee
Note that this expression is manifestly gauge invariant and have correct
commutative limit (\ref{e81}).

A generalization of this result to the case of $U(N)$ gauge group broken
down to $[U(1)]^N$ is rather trivial since the strength $W$ belonging to
$[U(1)]^N$ can be chosen as $W={\rm diag}(W_1,\ldots,W_N)$. Therefore the
effective action of such theory is a sum of actions (\ref{e84})
\be
\Gamma_{[U(1)]^N}[V^{++}]=\sum_{i=1}^N\Gamma_{U(1)}[V^{++}_i],
\label{e85}
\ee
where $\Gamma_{U(1)}[V^{++}_i]$ is given by eq. (\ref{e84}).

In the case of adjoint representation of gauge group the situation becomes
not so simple. Let us consider the classical action of $q$-hypermultiplet in
adjoint representation (\ref{e22}) when the vector superfield belongs to
the Cartan subalgebra of $u(N)$, i.e.
\be
Q^+=\sum\limits^N_{i,j=1}q^+_{ij}e_{ij},\qquad
V^{++}=\sum\limits^N_{k=1}V^{++}_ke_{kk},
\label{e86}
\ee
where
\be
(e_{ij})_{kl}=\delta_{ik}\delta_{jl},\qquad
i,j,k,l=1,\ldots,n
\label{basis}
\ee
is a Cartan-Weyl basis of $u(N)$ algebra \cite{Barut}. One can easy check
that the interaction (\ref{e22}) now reads
\be
\tr\int d\zeta^{(-4)}\breve Q^+\star[V^{++},Q^+]_\star
=\sum\limits^N_{k,l=1}\int d\zeta^{(-4)}
  (\breve q^+_{kl}\star V^{++}_k\star q^+_{kl}
 -\breve q^+_{kl}\star q^+_{kl}\star V^{++}_l).
\label{e87}
\ee
Therefore the contributions from hypermultiplets $q^+_{ij}$,
decouple and the formal expression for the 1-loop effective action reads
\be
\Gamma[V^{++}]=\sum\limits^N_{k,l=1}\Tr\ln\nabla^{++}_{kl},
\label{e88}
\ee
where the operators $\nabla^{++}_{kl}$ act on
hypermultiplets by the rule
$$
\nabla^{++}_{kl}q^+_{kl}={\cal D}^{++}q^+_{kl}+iV^{++}_k\star q^+_{kl}-
  iq^+_{kl}\star V^{++}_l.
$$
Each term in the sum (\ref{e88}) can be calculated in the same way as it was
done in sect. \ref{hyper}. A leading contribution to the effective action
is given by the term which is responsible for the UV/IR-mixing
\be
\Gamma[V^{++}]=\frac1{32\pi^2}\int d^4pd^8\theta
 du_1du_2\ln\frac4{m^2p\circ p}\frac{{\cal V}^{++}(p,\theta,u_1)
  {\cal V}^{++}(-p,\theta,u_2)}{(u^+_1u^+_2)^2},
\label{e90}
\ee
where ${\cal V}^{++}=\sum_k V^{++}_k$ is the gauge superfield related to the
$U(1)$ subgroup of $U(N)$ gauge group.
The expression (\ref{e90}) is singular in commutative limit.
We see that the UV/IR-mixing appear only in the $U(1)$ sector what was
firstly obtained for noncommutative $U(N)$ Yang-Mills model in \cite{Arm}
and for supersymmetric gauge theories in \cite{Armoni,Holl}.

The next order contribution to the effective action which is finite in
the limit $\theta\rightarrow0$ corresponds to the holomorphic potential of
commutative $q$-hypermultiplet in adjoint representation of $SU(N)$ group
calculated in \cite{Sams}.

To summarize, the effective action of noncommutative $q$-hypermultiplet in
adjoint representation of $U(N)$ group is singular in the commutative limit
and contains the terms which are responsible for the
UV/IR-mixing. The finite contribution in the limit $\theta\rightarrow0$
is given by the standard holomorphic potential of $q$-hypermultiplet in
adjoint representation of $SU(N)$ gauge group.

\subsection{Non-holomorphic potential in noncommutative ${\cal N}=4$ SYM}
Let us consider the ${\cal N}=4$ SYM model written in terms of ${\cal N}=2$ superfields
(\ref{e76}). The effective action of noncommutative ${\cal N}=4$ SYM (\ref{e77})
is determined by the functional integrals (\ref{e75}) in terms of noncommutative
covariant d'Alambertian (\ref{e73}). In commutative case it was shown in
refs. \cite{Buch3,Buch4} that the effective action can be represented in
the form of functional integral over unconstrained ${\cal N}=1$ superfields
what allows to restore the non-holomorphic potential very
easy. We will see that the effective action of noncommutative ${\cal N}=4$
SYM can also be represented as a functional integral over unconstrained
${\cal N}=1$ superfields what gives a starting point for calculating the
noncommutative corrections to the holomorphic potential. This is done
only with the help of special constraints on the background strength superfield $W$.
The first requirement is the on-shell constraint
\be
{\cal D}^{\alpha(i}\star{\cal D}^{j)}_\alpha\star W=0
\label{e92}
\ee
that simplifies the noncommutative covariant d'Ala\-mber\-ti\-an to the form
\be
\hBox_\star={\cal D}^m\star{\cal D}_m+\frac12\{\bar W,W\}_\star
+\frac i2(D^{+\alpha}W)\star{\cal D}^-_\alpha+
\frac i2(\bar D^+_{\dot\alpha}\bar W)\star\bar{\cal D}^{-\dot\alpha}.
\label{e93}
\ee
Moreover, we consider only the case when the gauge symmetry $U(N)$ is
broken down to $[U(1)]^N$ and the strength $W$ belongs to the Cartan
subalgebra of $u(N)$. Further calculations are very similar to ones made
for conventional ${\cal N}=4$ SYM in harmonic superspace given in refs.
\cite{Buch3,Buch4}.

At first, we make a replacement of functional variables in the integrals
(\ref{e75})
\be
\begin{array}l
v^{++}={\cal F}^{++}+\nabla^{++}\star\sigma\\
u^{++}={\cal G}^{++}+\nabla^{++}\star\dint d\tilde\zeta^{(-4)}
 G^{(0,0)}(\zeta,\tilde\zeta)\rho^{(+4)}(\tilde\zeta),
\end{array}
\label{e94}
\ee
where ${\cal F}^{++}$ and ${\cal G}^{++}$ are constrained ${\cal N}=2$
superfields
\be
\nabla^{++}\star{\cal F}^{++}=0,\qquad \nabla^{++}\star G^{++}=0.
\label{e96}
\ee
This allows us to rewrite the effective action in the form of single
functional integral but over the constrained superfield ${\cal F}^{++}$:
\be
\exp({i\Gamma^{(1)}_{{\cal N}=4}})=\tfrac{
\dint{\cal DF}^{++}\exp\{-\tfrac i2\tr\dint d\zeta^{(-4)}
  {\cal F}^{++}\star\hBox_\star\star{\cal F}^{++} \}
}{
\dint{\cal DF}^{++}\exp\{\tfrac i2\tr\dint d\zeta^{(-4)}
  {\cal F}^{++}{\cal F}^{++} \}
}.
\label{e97}
\ee
The superfield ${\cal F}^{++}$ can be expanded over harmonic variables as
\be
{\cal F}^{++}(z,u)={\cal F}^{ij}(z)u^+_iu^+_j,
\label{e98}
\ee
where the superfields ${\cal F}^{ij}$ satisfy the constraints
\be
{\cal D}^{(i}_\alpha\star{\cal F}^{jk)}=\bar{\cal D}^{(i}_{\dot\alpha}
 \star{\cal F}^{jk)}=0,\qquad \bar{\cal F}^{ij}={\cal F}_{ij}.
\label{e99}
\ee
The operator $\hBox_\star$ acts on ${\cal F}^{ij}$ as follows
\be
\hBox_\star{\cal F}^{ij}=(
{\cal D}^m\star{\cal D}_m+\frac12\{\bar W,W\}_\star)\star{\cal F}^{ij}
+
\frac i3({\cal D}^{\alpha(i}\star W)\star{\cal D}_{\alpha|k|}\star
 {\cal F}^{j)k} +
\frac i3(\bar D_{\dot\alpha}^{(i}\star\bar W)
 \star\bar{\cal D}^{\dot\alpha}_{|k|}\star{\cal F}^{j)k}.
\label{e100}
\ee

The next step is to go to ${\cal N}=1$ projections of ${\cal N}=2$ superfield
${\cal F}^{++}$ in order to reduce the functional integral (\ref{e97}) to
one over unconstrained superfields. We introduce ${\cal N}=1$ Grassman
coordinates $(\theta^\alpha,\bar\theta_{\dot\alpha})$ by the rule
$\theta^\alpha=\theta^\alpha_1$,
$\bar\theta_{\dot\alpha}=\bar\theta^1_{\dot\alpha}$, the corresponding
gauge covariant derivatives ${\cal D}_\alpha={\cal D}^1_\alpha$,
$\bar{\cal D}^{\dot\alpha}=\bar{\cal D}^{\dot\alpha}_1$ and then define
the ${\cal N}=1$ projections of an arbitrary ${\cal N}=2$ superfield as
$f|=f(x^m,\theta^\alpha_i,\bar\theta^i_{\dot\alpha})|_{
\theta_2=\bar\theta^2=0}$. The ${\cal N}=1$ projections of ${\cal N}=2$ strength $W$ and
${\cal F}^{ij}$ read
\be
\begin{array}l
\phi=W|,\qquad 2iW_\alpha={\cal D}_\alpha^2\star W|,\\
\Psi={\cal F}^{22}|,\qquad \bar\Psi={\cal F}^{11}|,\qquad
F=\bar F=-2i{\cal F}^{12}|.
\end{array}
\label{e101}
\ee
The equations (\ref{e99}) reduce to the following constraints on $\Psi,F$
\be
\bar{\cal D}_{\dot\alpha}\star\Psi=0,\qquad
-\frac14(\bar{\cal D})^2\star F+[\phi,\Psi]_\star=0.
\label{e102}
\ee

At this point we require the ${\cal N}=1$ components of background strength
superfield $W$ to be covariantly constant
\be
{\cal D}_\alpha\star\phi=0,\qquad
 {\cal D}_\alpha\star W_\beta=0.
\label{e103}
\ee
In commutative case such constraints were sufficient to calculate a
non-holomorphic potential. We are interested in noncommutative corrections
to this result, therefore we also require such constraints in
noncommutative case. For such a background the operator $\hBox_\star$ does
not mix the superfields $\Psi,\bar\Psi$ and $F$
\be
(\hBox_\star\star{\cal F}^{ij})|=\Delta\star({\cal F}^{ij}),
\label{e104}
\ee
where
\be
\Delta={\cal D}^m\star{\cal D}_m+\frac12\{\phi,\bar\phi\}_\star-
 W^\alpha\star{\cal D}_\alpha+
  \bar W_{\dot\alpha}\star\bar{\cal D}^{\dot\alpha}.
\label{e105}
\ee
This allows us to rewrite the functional integral (\ref{e97}) in the form
\be
\exp({i\Gamma^{(1)}_{{\cal N}=4}})=\tfrac{
\dint{\cal D}\bar\Psi{\cal D}\Psi{\cal D}F\exp\{
i\tr\int d^8z(-\bar\Psi\star\Delta\star\Psi+\tfrac12 F\star\Delta\star F)\}
}{
\dint{\cal D}\bar\Psi{\cal D}\Psi{\cal D}F\exp\{
i\tr\int d^8z(-\bar\Psi\Psi+\tfrac12 F^2)\}
}.
\label{e106}
\ee

The superfields $\Psi,\bar\Psi,F$ in the expression (\ref{e106}) are the
Lie algebra valued superfields of $U(N)$ gauge group while the background
strength $W_\alpha$ and $\phi$ belong to the Cartan subalgebra. They
can be written in the Cartan-Weyl basis of $U(N)$ (\ref{basis}) as
\be
F=\tsum\limits^N_{k\ne l}F^{kl}e_{kl}+\tsum\limits^N_{k=1}F^ke_{kk},\quad
\phi=\tsum\limits^N_{k=1}\phi^ke_{kk}.
\label{e107}
\ee
Given an element $\phi$ in the Cartan subalgebra, one finds
\be
[\phi,F]_\star=\sum\limits^N_{k\ne l}(\phi^k\star F^{kl}-
 F^{kl}\star\phi^l)e_{kl}+
\sum\limits^N_{k=1}(\phi^k\star F^k-F^k\star\phi^k)e_{kk}.
\ee
Therefore, the second equation of constraints (\ref{e102}) can be written as
\be
\begin{array}l
-\frac14(\bar{\cal D}^2\star F^{kl})+(\phi^k\star\Psi^{kl}-
  \Psi^{kl}\star\phi^l)=0,\\
-\frac14(\bar{\cal D}^2\star F^k)+(\phi^k\star\Psi^k-\Psi^k\star\phi^k)=0.
\end{array}
\label{e109}
\ee
The equations (\ref{e109}) can be resolved with respect to the superfields
$\Psi^{kl}$ and $\Psi^k$ as a series in $\theta$~\footnote{
A solution to the equation in the second line of (\ref{e109})
is not unique since in the limit $\theta\rightarrow0$ this eq. becomes the
identity, but it does not spoil the picture.}. A formal solution is
written as
\be
\Psi^{kl}={\bf B}^{kl}F^{kl},\qquad \psi^k={\bf B}^kF^k,
\label{e110}
\ee
where ${\bf B}^{kl}$, ${\bf B}^k$ are some operators which we do not
specify.
Taking into account the eqs. (\ref{e110}) one can transform the integral
in the denominator of eq. (\ref{e106}) as follows
\be
\begin{array}{l}
\tr\dint d^8z(-\bar\Psi\star\Psi+\frac12F\star F)\\\qquad=
 \dint d^8z\sum\limits^N_{k\ne l}(\bar {\bf B}^{kl}F^{kl}\star{\bf B}^{kl}F^{kl}+
  \frac12F^{kl}\star F^{kl})
+ \dint d^8z\sum\limits^N_{k=1}({\bf B}^kF^k\star{\bf B}^kF^k+
 \frac12F^k\star F^k).
\end{array}
\label{e111}
\ee
As a result, we obtain the following formal expression for the
functional integral in the denominator of eq. (\ref{e106})
\be
\begin{array}l
\dint{\cal D}\bar\Psi{\cal D}\Psi{\cal D}F\exp\{
i\tr\dint d^8z(-\bar\Psi\star\Psi+\frac12F\star F)\}\\
\qquad=\dint{\cal D}\bar V^{kl}{\cal D}V^{kl}{\cal D}V^k\exp\{
i\int d^8z(\sum\limits_{k<l}^N\bar V^{kl}{\bf C}_{kl}V^{kl}
+\sum\limits_{k=1}^NV^k{\bf C}_kV^k)\}\\
\qquad=\prod\limits_{k<l}^N{\rm Det}^{-1}({\bf C}_{kl})
\prod\limits_{k=1}^N{\rm Det}^{-1}({\bf C}_k),
\end{array}
\label{e112}
\ee
where we introduced the superfields $V^{kl}=F^{kl}$, $\bar V^{kl}=F^{lk}$,
$k<l$, and the operators ${\bf C}_{kl}$, ${\bf C}_k$ which act as
\be
\begin{array}l
\dint d^8z \bar{\bf B}^{kl}F^{kl}\star{\bf B}^{kl}F^{kl}=
 \dint d^8z F^{kl}\star{\bf C}^{kl}F^{kl},\\[2mm]
\dint d^8z {\bf B}^{k}F^{k}\star{\bf B}^{k}F^{k}=
 \dint d^8z F^{k}\star{\bf C}^{k}F^{k}.
\end{array}
\label{e113}
\ee

Following the same steps we represent the numerator of eq. (\ref{e106})
in the form
\be
\begin{array}l
\dint{\cal D}\bar\Psi{\cal D}\Psi{\cal D}F\exp\{
i\tr\int d^8z(-\bar\Psi\star\Delta\star\Psi
 +\frac12F\star\Delta\star F)\}\\
\qquad=\int{\cal D}\bar V^{kl}{\cal D}V^{kl}{\cal D}V^k\exp\{
i\int d^8z(\sum\limits_{k<l}^N\bar V^{kl}
 \star{\bf C}_{kl}\Delta_{kl}\star V^{kl}+
\sum\limits_{k=1}^NV^k\star{\bf C}_k\Delta_k\star V^k)\}\\
\qquad=\prod\limits_{k<l}^N{\rm Det}^{-1}({\bf C}_{kl})
\prod\limits_{k=1}^N{\rm Det}^{-1}({\bf C}_k)
\prod\limits_{k<l}^N{\rm Det}^{-1}(\Delta_{kl})
\prod\limits_{k=1}^N{\rm Det}^{-1}(\Delta_k),
\end{array}
\label{e114}
\ee
where the operators $\Delta_{kl}$, $\Delta_k$ act on the superfields
$V^{kl}$, $V^{k}$ by the rules:
\be
\begin{array}{l}
\Delta_{kl}\star V^{kl}={\cal D}^m\star{\cal D}_m\star V^{kl}\\\quad
 -(W^{k\alpha}\star({\cal D}_\alpha\star V^{kl})-
 ({\cal D}_\alpha\star V^{kl})\star W^{l\alpha})
+(\bar W^k_{\dot\alpha}\star(\bar{\cal D}^{\dot\alpha}\star V^{kl})-
 (\bar{\cal D}^{\dot\alpha}\star V^{kl})\star\bar W^l_{\dot\alpha})\\
\quad+\bar\phi^k\star\phi^k\star V^{kl}-2\bar\phi^k\star V^{kl}\star\phi^l
-2\phi^k\star V^{kl}\star\bar\phi^l\\
\quad+V^{kl}\star\phi^l\star\bar\phi^l
+ \phi^k\star\bar\phi^k\star V^{kl}+V^{kl}\star\bar\phi^l\star\phi^l,
\end{array}
\label{e115}
\ee
\be
\begin{array}{l}
\Delta_{k}\star V^{k}={\cal D}^m\star{\cal D}_m\star V^{k}\\\quad-
 (W^{k\alpha}\star({\cal D}_\alpha\star V^{k})-
 ({\cal D}_\alpha\star V^{k})\star W^{k\alpha})
+(\bar W^k_{\dot\alpha}\star(\bar{\cal D}^{\dot\alpha}\star V^{k})-
 (\bar{\cal D}^{\dot\alpha}\star V^{k})\star\bar W^k_{\dot\alpha})\\
\quad+\bar\phi^k\star\phi^k\star V^{k}-2\bar\phi^k\star V^{k}\star\phi^k
-2\phi^k\star V^{k}\star\bar\phi^k+\\
\quad V^{k}\star\phi^k\star\bar\phi^k
+\phi^k\star\bar\phi^k\star V^{k}+V^{k}\star\bar\phi^k\star\phi^k.
\end{array}
\label{e116}
\ee
Substituting the functional integrals (\ref{e112},\ref{e114}) into eq.
(\ref{e106}) we see that the operators ${\bf C}^k$ and ${\bf C}^{kl}$
in numerator and denominator cancel each other
 and the resulting expression for the effective action
is represented in the following form
\be
\begin{array}c
\Gamma^{(1)}_{{\cal N}=4}=\sum\limits_{k<l}^N\Gamma_{kl}+
 \sum\limits^N_{k=1}\Gamma_k,\\
\Gamma_{kl}=i\Tr\ln\Delta_{kl},\qquad \Gamma_k=i\Tr\ln\Delta_k.
\end{array}
\label{e117}
\ee
The formal expressions of $\Tr\ln$ of the operators $\Delta_{kl}$,
$\Delta_k$ should be understood in the sense of the functional integrals
over the ${\cal N}=1$ superfields $V^{kl}$, $V^{k}$
\be
\begin{array}l
{\rm Det}^{-1}(\Delta_{kl})=
\dint{\cal D}\bar V^{kl}{\cal D}V^{kl}\exp\{
i\dint d^8z(\bar V^{kl}\star\Delta_{kl}\star V^{kl}\},\\
{\rm Det}^{-1/2}(\Delta_k)
=\dint{\cal D}V^k\exp\{
i\dint d^8z(V^k\star\Delta_k\star V^k)\}.
\end{array}
\label{e118}
\ee
The first term $\sum_{k<l}^N\Gamma_{kl}$ corresponds to
$U(N)/[U(1)]^N$ sector of $U(N)$ group while the second term is defined by
the integrals over superfields belonging to the $[U(1)]^N$ subgroup.

It should be noted that the commutative limit is smooth since ${\cal N}=4$ SYM
has no divergencies. Taking the limit $\theta\rightarrow0$ in
the expressions (\ref{e115},\ref{e116}) we see that the operator
$\Delta_{kl}$ reproduces the standard non-holomorphic potential of
commutative ${\cal N}=4$ $SU(N)$ SYM studied in \cite{Buch4},
while the contributions from the operator $\Delta_k$ vanish. Therefore, the
second term of effective action $\sum_{k=1}^N\Gamma_k$ is responsible
for higher noncommutative corrections only.

Thus, we have obtained here a useful representation (\ref{e117})
of 1-loop effective action of
noncommutative ${\cal N}=4$ SYM in terms of functional integral over
unconstrained ${\cal N}=1$ superfields what provides a basis for perturbative
calculations of 1-loop effective action and studying the noncommutative
corrections.

\section{Summary}
We have developed a formulation of the basic ${\cal N}=2$ supersymmetric
field models in noncommutative harmonic superspace. The classical actions of
noncommutative hypermultiplet and ${\cal N}=2$ vector multiplet can be easily
constructed in such a superspace in terms of unconstrained ${\cal N}=2$
superfields. We formulated the Feynman rules for these theories,
considered the calculations of various one-loop harmonic supergraphs
and pointed out the new aspects arising in comparison with
corresponding commutative theories.

We observed that the ${\cal N}=2$ SYM model has a standard (for
noncommutative theories) UV/IR-mixing and its effective action is singular
in the limit of small noncommutativity $\theta\rightarrow0$. The
model of Abelian noncommutative $q$-hypermultiplet interacting with
external vector superfield is free of UV/IR-mixing and has no non-planar
diagram contributions. The model of selfinteracting $q$-hypermultiplet is
nonrenormalizable but the quartic interaction vertex can be induced by
1-loop quantum corrections as it is in the commutative case.

We formulated a background field method in noncommutative harmonic
superspace and considered a structure of low-energy effective action of
general noncommutative ${\cal N}=2$ model. It is shown that the one-loop effective
action is represented in terms of superfield functional determinants of the
operators $\hBox_\star$ and $\nabla^{++}$ described in the paper.
We found that the holomorphic
effective action of noncommutative $q$-hypermultiplet in the fundamental
representation is free of non-planar contributions
and it can be written in manifestly supersymmetric and
gauge invariant form as in commutative case. The effective action of
$q$-hypermultiplet in adjoint representation has UV/IR-mixing due to the
nonplanar contributions what does not allow us to represent the effective
action in a gauge invariant form.

We have considered the one-loop effective action of noncommutative
${\cal N}=4$ SYM theory. This theory has been studied within the
${\cal N}=2$ background
field method and it has been shown that the effective action consists of
two contributions associated with $U(N)/U(1)^N$ and $U(1)^N$ sectors of the
$U(N)$ gauge group.  The first
term reproduces a standard non-holomorphic potential in
$\theta\rightarrow0$ limit while the second one is responsible for higher
noncommutative corrections only.

The constructions developed in the paper and the results concerning the
formulation of ${\cal N}=2$ background field method in noncommutative
harmonic superspace and the structure of low-energy effective action can
be applied for the study of various quantum aspects in noncommutative
${\cal N}=2$ supersymmetric field models.\\[1cm]
{\bf Acknowledgements}\\
The authors would like to thank B.M. Zupnik for valuable
discussions and A. Armoni for useful comments.
I.L.B. is very grateful to S.J.~Gates for the kind
hospitality in the University of Maryland during the final stage of
this work. The work
was supported in part by INTAS grant, INTAS-00-00254. The work of I.L.B.
was also supported partially by RFBR grant No 99-02-16617 and
INTAS grant, INTAS-99-1590.


\end{document}